\documentclass[reprint,floatfix,amsmath,amssymb,superscriptaddress]{revtex4-1}

\usepackage{graphicx}
\usepackage{bm}
\usepackage{epstopdf}
\usepackage{xr}
\usepackage{float}
\usepackage{lineno}
\usepackage{siunitx}
\usepackage{xcolor}
\usepackage{cleveref}
\usepackage{soul}

\begin{document}

\title{Comments on the formula to extract current-induced torques from the harmonic Hall voltage measurements}

\author{Yong-Chang Lau}
\affiliation{Department of Physics, The University of Tokyo, Bunkyo-ku, Tokyo 113-0033, Japan}

\author{Yukihiro Marui}
\affiliation{Research Institute of Electrical Communication (RIEC), Tohoku University, Sendai 980-8577, Japan}

\author{Zhendong Chi}
\affiliation{Department of Physics, The University of Tokyo, Bunkyo-ku, Tokyo 113-0033, Japan}

\author{Masashi Kawaguchi}
\affiliation{Department of Physics, The University of Tokyo, Bunkyo-ku, Tokyo 113-0033, Japan}

\author{Masamitsu Hayashi}
\affiliation{Department of Physics, The University of Tokyo, Bunkyo-ku, Tokyo 113-0033, Japan}
\affiliation{Trans-Scale Quantum Science Institute (TSQS), The University of Tokyo,
Tokyo 113-0033, Japan}
\date{\today}



\begin{abstract}
We examine the formulas commonly used to estimate current-induced spin-orbit torques from harmonic Hall voltage measurements. In particular, we focus on the factor of two discrepancy among expressions employed to fit harmonic Hall signals measured under an in-plane rotating magnetic field. By explicitly deriving the relevant relations, we clarify the origin of this discrepancy and present the correct form of the fitting formula. We further discuss the determination of the sign of the field-like torque from harmonic Hall voltage measurements, which depends on the assumed form of the current-induced torques.
\end{abstract}

\maketitle
\section{Introduction}
The harmonic Hall voltage measurements has been used to determine current-induced torque in metallic bilayers consisting of a nonmagnetic metal (NM) and a ferromagnetic metal (FM).
The measurement scheme can be categorized into two classes.
The original approach\cite{pi2010apl,kim2013nmat,garello2013nnano,hayashi2014prb} was applied to NM/FM bilayers with the FM layer having an out of plane magnetic easy axis. 
The external magnetic field was swept along the film plane, parallel and orthogonal to the current flow direction to estimate the damping-like and field-like components of the current-induced torque.
Here the magnetic field must be kept small enough so that the FM layer magnetization remains close to its equilibrium direction.
The second approach was applied to bilayers with the FM layer having an in plane magnetic easy axis\cite{avci2014prb}.
The direction of the external magnetic field is rotated within the film plane and the angular dependence of the harmonic Hall voltage is fitted with sinusoidal functions to estimate the current-induced torque.
To extract the current-induced torque accurately, the applied magnetic field is typically larger than that of the first approach\cite{avci2014prb,roschewsky2019prb}.
Hereafter, we refer to the second approach as the rotating-field harmonic Hall voltage measurements.

We find that the fitting function for the rotating-field measurements is off by a factor of 2 in some publications\cite{avci2014prb,lau2017jjap,chi2020sciadv,chi2021aplmater}.
To clarify this, we first derive the correct form of the fitting function.
We follow the model described in Ref.~\cite{hayashi2014prb}.
Although Ref.~\cite{hayashi2014prb} did not discuss the rotating-field measurements, the formula provided can be used for the derivation.
The correct form of the harmonic Hall resistance is provided in Eq.~(\ref{eq:R2w}) (one must be aware of how the parameters are defined).

\section{Model description}
\subsection{Magnetic system}
The magnetization $\bm{M}$ of the FM layer is denoted as
\begin{equation}
\begin{aligned}
\label{eq:M}
\bm{M} = M_\mathrm{S} \bm{m} = M_\mathrm{S} (\sin\theta \cos\varphi, \sin\theta \sin\varphi, \cos\theta),
\end{aligned}
\end{equation}
where $\bm{m}$ is an unit vector.
The magnetic energy density of the system reads
\begin{equation}
\begin{aligned}
\label{eq:energy}
E = - K_\mathrm{EFF} \cos^2\theta - K_\mathrm{I} \sin^2 \theta \sin^2 \varphi- \bm{M} \cdot \bm{H} ,
\end{aligned}
\end{equation}
where $K_\mathrm{EFF}$ is the effective out of plane magnetic anisotropy energy density, $K_\mathrm{I}$ is the in-plane uniaxial magnetic anisotropy energy density and $\bm{H}$ is the external magnetic field, which is defined as 
\begin{equation}
\begin{aligned}
\label{eq:H}
\bm{H} = H (\sin\theta_H \cos\varphi_H, \sin\theta_H \sin\varphi_H, \cos\theta_H).
\end{aligned}
\end{equation}
We define the effective out of plane anisotropy field $H_K$ and the in-plane uniaxial magnetic anisotropy field $H_\mathrm{A}$ as 
\begin{equation}
\begin{aligned}
\label{eq:HK}
H_K = \frac{2 K_\mathrm{EFF}}{M_\mathrm{S}}, \ H_\mathrm{A} = \frac{2 K_\mathrm{I}}{M_\mathrm{S}}.
\end{aligned}
\end{equation}
$H_K$ is positive (negative) when the magnetic easy axis of the FM layer points along the film normal (film plane).
We set $H_\mathrm{A} = 0$ hereafter since $H_\mathrm{A}$ is typically significantly smaller than the external magnetic field $H$.

The equilibrium magnetization state can be obtained by solving the following equations:
\begin{equation}
\begin{aligned}
\label{eq:equil}
\frac{\partial E}{\partial \theta} = 0, \ \frac{\partial E}{\partial \varphi} = 0.
\end{aligned}
\end{equation}
Substituting Eq.~(\ref{eq:energy}) into Eq.~(\ref{eq:equil}) returns the equilibrium magnetization $\bm{m}_0$, which we denote as 
\begin{equation}
\begin{aligned}
\label{eq:m0}
\hat{\bm{m}}_0 = (\sin\theta_0 \cos\varphi_0, \sin\theta_0 \sin\varphi_0, \cos\theta_0).
\end{aligned}
\end{equation}
The solution of the second equation of Eq.~(\ref{eq:equil}) gives 
\begin{equation}
\begin{aligned}
\label{eq:varphi}
\varphi_0 = \varphi_H.
\end{aligned}
\end{equation}

\subsection{current-induced torque}
We now consider the influence of current-induced torque on the magnetization when current is applied to the bilayer.
The effective field associated with the current-induced torque is defined as $\Delta \bm{H}$.
We assume that the torque consists of two components: damping-like and field-like torques.
$\Delta \bm{H}$ is given by
\begin{equation}
\begin{aligned}
\label{eq:Heff}
\Delta \bm{H} = \frac{1}{M_\mathrm{s} t} \left( \hat{\bm{m}}_0 \times \bm{p} + \beta \bm{p} \right),
\end{aligned}
\end{equation}
where the first (second) term on the right hand side is the damping-like (field-like) component.
$\bm{p}$ is the spin current that enters the FM layer.
The direction and magnitude of $\bm{p}$ represent the polarization and size of the spin current, respectively.
Here the polarization of spin current corresponds to the direction of the spin magnetic moment (not the spin angular momentum) of the carriers. 
$\beta$ is a coefficient that characterizes the field-like torque with respect to the damping-like torque and $t$ is the thickness of the FM layer. 

As a general example, let us consider the case when the spin current from the NM layer, generated via the spin Hall effect, exerts spin torque on the magnetization of the FM layer.
If the FM layer is placed on top of (beneath) the NM layer, the spin current that flows into the FM layer is along $+z$ ($-z$).
$\bm{p}$ is therefore expressed as 
\begin{equation}
\begin{aligned}
\label{eq:js:pol}
\bm{p} = T \theta_\mathrm{SH} \frac{\hbar}{2e} \left( \bm{j}_\mathrm{c} \times \left( \eta \hat{\bm{e}}_z \right) \right)
\end{aligned}
\end{equation}
where $\theta_\mathrm{SH}$ is the spin Hall angle of the NM layer and $T$ is the interface spin transmission probability, which is often characterized using the spin mixing conductance\cite{weiler2013prl,kim2014prb}.
$\hat{\bm{e}}_i$ is an unit vector along the $i$-direction.
$\eta$ represents the film stacking: $+1$ for sub./NM/FM (spin current flows along $+z$) and $-1$ for sub./FM/NM (spin current flows along $-z$).
Let us assume that current flows along $+x$, i.e. $\bm{j}_\mathrm{c} = j_\mathrm{c} \hat{\bm{e}}_x$.
We therefore have 
\begin{equation}
\begin{aligned}
\label{eq:js:pol:y}
\bm{p} &= T \theta_\mathrm{SH} \frac{\hbar}{2e} j_\mathrm{c} (\hat{\bm{e}}_x \times \left( \eta \hat{\bm{e}}_z \right) ) = -\eta T \theta_\mathrm{SH} \frac{\hbar}{2e} j_\mathrm{c} \hat{\bm{e}}_y
\end{aligned}
\end{equation}
Substituting Eq.~(\ref{eq:js:pol:y}) into Eq.~(\ref{eq:Heff}), we obtain 
\begin{equation}
\begin{aligned}
\label{eq:sot}
\Delta \bm{H} &= - \eta \frac{\hbar T \theta_\mathrm{SH} j_\mathrm{c}}{2 e M_\mathrm{s} t} \left( \hat{\bm{m}}_0 \times \hat{\bm{e}}_y +\beta \hat{\bm{e}}_y \right)\\
& = - h_\mathrm{DL} \left( \hat{\bm{m}}_0 \times \hat{\bm{e}}_y +\beta \hat{\bm{e}}_y \right),
\end{aligned}
\end{equation}
where we defined the magnitude of the damping-like effective field as
\begin{equation}
\begin{aligned}
\label{eq:sot:aJbJ}
h_\mathrm{DL} = \eta \frac{\hbar T \theta_\mathrm{SH} j_\mathrm{c}}{2 e M_\mathrm{s} t}.
\end{aligned}
\end{equation}
In components, we have
\begin{equation}
\begin{aligned}
\label{eq:sot:aJbJ}
(\Delta H_x,& \Delta H_y, \Delta H_z)\\
&= (h_\mathrm{DL} \cos\theta_0, - \beta h_\mathrm{DL}, -h_\mathrm{DL} \sin\theta_0 \cos\varphi_0).
\end{aligned}
\end{equation}
We will later use Eq.~(\ref{eq:sot:aJbJ}) to obtain the form of harmonic Hall voltages [Eq.~(\ref{eq:sot:aJbJ}) is substituted into Eq.~(\ref{eq:V2w:inp})].

\subsection{Harmonic Hall votlages}
Let us assume that the current is small enough such that the magnetization slightly tilts from its equilibrium direction.
We define the change in the magnetization angle induced by the current as $\Delta \theta$ and $\Delta \varphi$.
Following the approach described in Ref.~\cite{hayashi2014prb}, $\Delta \theta$ and $\Delta \varphi$ read
\begin{equation}
\begin{aligned}
\label{eq:deltatheta}
\Delta \theta = \frac{\cos\theta_0 (\Delta H_x \cos\varphi_H + \Delta H_y \sin\varphi_H ) - \sin\theta_0 \Delta H_z }{H_K \cos 2\theta_0 + H \cos(\theta_H - \theta_0) },\\
\end{aligned}
\end{equation}
\begin{equation}
\begin{aligned}
\label{eq:deltavarphi}
\Delta \varphi = \frac{- \Delta H_x \sin\varphi_H + \Delta H_y \cos\varphi_H }{ H \sin\theta_H }.
\end{aligned}
\end{equation}
Note that, under $H_\mathrm{A} = 0$, Eqs.~(\ref{eq:deltatheta}) and (\ref{eq:deltavarphi}) are exact, that is, there is no assumption made in deriving these relations. This is because we dropped the in-plane uniaxial magnetic anisotropy energy ($K_\mathrm{I} = 0$), which simplified the solutions.
Equations~(\ref{eq:deltatheta}) and (\ref{eq:deltavarphi}) are equivalent to, respectively, Eqs.~(14) and (15) in Ref.~\cite{hayashi2014prb} with $H_\mathrm{A} = 0$.

With $\Delta \theta$ and $\Delta \varphi$, we estimate the change in the Hall resistance due to current-induced torque. First, the Hall resistance $R_{yx}$ due to the anomalous Hall and planar Hall effects are given as
\begin{equation}
\begin{aligned}
\label{eq:Ryx}
R_{yx} = a \left( R_\mathrm{A} \cos\theta + R_\mathrm{P} \sin^2 \theta \sin 2 \varphi \right),
\end{aligned}
\end{equation}
where $R_\mathrm{A}$ and $R_\mathrm{P}$ are the amplitudes of the anomalous Hall and planar Hall resistances, respectively.
$a$ is a constant that depends on how one defines the amplitude of the Hall resistances.
In Ref.~\cite{hayashi2014prb}, $a = \frac{1}{2}$ whereas Ref.~\cite{avci2014prb} used $a = 1$.
Later on, we substitute $a = 1$ to compare the results with the literature.

We consider the case when a small current is applied to the bilayer.
The current-induced torque tilts the magnetization away from its equilibrium direction.
We therefore substitute 
\begin{equation}
\begin{aligned}
\label{eq:angle:current}
\theta = \theta_0 + \Delta \theta,\ \varphi = \varphi_0 + \Delta \varphi
\end{aligned}
\end{equation}
into Eq.~(\ref{eq:Ryx}) and use first order Taylor expansion with $\Delta \theta \ll 1$ and $\Delta \varphi \ll 1$, which give
\begin{equation}
\begin{aligned}
\label{eq:Ryx:small}
R_{yx} = & a R_\mathrm{A} \left( \cos\theta_0 - \Delta \theta \sin\theta_0 \right)\\
&+ a R_\mathrm{P} \left( \sin^2 \theta_0 + \Delta \theta \sin 2 \theta_0 \right) \left( \sin 2 \varphi_0 + 2 \Delta \varphi \cos 2 \varphi_0 \right).
\end{aligned}
\end{equation}
Eq.~(\ref{eq:Ryx:small}) represents the Hall resistance when a small current is applied to the bilayer. 

In the harmonic Hall voltage measurements, the current applied to the bilayer is an ac current.
The current $I$ is defined as 
\begin{equation}
\begin{aligned}
\label{eq:I}
I = \Delta I \sin \omega t,
\end{aligned}
\end{equation}
where $\Delta I$ and $\omega$ are the amplitude and the angular frequency of the sinusoidal current $I$.
The Hall voltage therefore reads
\begin{equation}
\begin{aligned}
\label{eq:Vyx}
V_{yx} = R_{yx} I = V_0 + V_\omega \sin \omega t + V_{2 \omega} \cos 2 \omega t,
\end{aligned}
\end{equation}
where the rectified dc voltage ($V_0$), the first ($V_\omega$) and 90 deg out-of-phase second ($V_{2 \omega}$) harmonic voltages are given as 
\begin{equation}
\begin{aligned}
\begin{split}
\label{eq:V2w}
V_0 &= - V_{2 \omega},\\
V_\omega &= \Delta I a \left(R_\mathrm{A} \cos\theta_0 + R_\mathrm{P} \sin^2 \theta_0 \sin 2 \varphi_0 \right),\\
V_{2 \omega} &= -\frac{1}{2} \Delta I a \left[ \left( - R_\mathrm{A} \sin\theta_0 + R_\mathrm{P} \sin 2 \theta_0 \sin 2 \varphi_0 \right) \Delta \theta \right.\\
&\ \ \ \ \ \ \ \ \ \ \ \ \ \ \ \ \left. + \left( 2 R_\mathrm{P}\sin^2\theta_0 \cos 2 \varphi_0 \right) \Delta \varphi \right] 
\end{split}
\end{aligned}
\end{equation}
Equations~(\ref{eq:Vyx}) and (\ref{eq:V2w}) are the same with Eq.~(19) of Ref.~\cite{hayashi2014prb} ($a = \frac{1}{2}$ in Ref.~\cite{hayashi2014prb}).

From hereon, we consider a case when a large in-plane magnetic field is applied to the bilayer such that the magnetization direction of the FM layer follows the magnetic field.
This is the experimental configuration used for the rotating-field harmonic Hall voltage measurements.
We therefore set
\begin{equation}
\begin{aligned}
\label{eq:config:inp}
\theta_0 = \theta_H = \frac{\pi}{2}, \ \varphi_0 = \varphi_H.
\end{aligned}
\end{equation}
The latter relation derives from Eq.~(\ref{eq:varphi}).
Substituting these relations and Eqs.~(\ref{eq:deltatheta}), (\ref{eq:deltavarphi}) into Eq.~(\ref{eq:V2w}), we find
\begin{equation}
\begin{aligned}
\begin{split}
\label{eq:V2w:inp}
V_\omega &= \Delta I a R_\mathrm{P} \sin 2 \varphi_{H},\\
V_{2 \omega} &= -\frac{1}{2} \Delta I a \left[ R_\mathrm{A} \frac{ \Delta H_z }{H - H_K} \right.\\
& \ \ \ \ \ \ \ \left. +  2 R_\mathrm{P} \frac{ \cos 2 \varphi_H \left( - \Delta H_x \sin\varphi_H + \Delta H_y \cos\varphi_H \right) }{H} \right] 
\end{split}
\end{aligned}
\end{equation}
We use Eq.~(\ref{eq:sot:aJbJ}) to express the current-induced torque $\Delta \bm{H} = (\Delta H_x, \Delta H_y, \Delta H_z)$ with the damping-like ($h_\mathrm{DL}$) and field-like ($\beta h_\mathrm{DL}$) components.
Substituting Eq.~(\ref{eq:sot:aJbJ}) into the second line of Eq.~(\ref{eq:V2w:inp}), we obtain
\begin{equation}
\begin{aligned}
\begin{split}
\label{eq:V2w:inp:aJbJ}
V_{2 \omega} =& \frac{1}{2} \Delta I a \left[ R_\mathrm{A} \frac{ h_\mathrm{DL} }{H - H_K} \cos\varphi_H \right.\\
& \ \ \ \ \ \ \ \ \ \ \ \ \ \ \  \left. + 2 R_\mathrm{P} \frac{ \beta h_\mathrm{DL} }{H} \cos\varphi_H \cos 2 \varphi_H \right] 
\end{split}
\end{aligned}
\end{equation}
Equation~(\ref{eq:V2w:inp:aJbJ}) is one of the main results of this paper. As the form can vary depending on the definition of material parameters, we discuss these in the following.

\section{Discussion}
\subsection{Sign of the field-like torque}
First, it should be noted that the current-induced torque can be expressed using different forms. 
For example, one may rewrite Eq.~(\ref{eq:Heff}) as the following:
\begin{equation}
\begin{aligned}
\label{eq:sot:pcrossm}
\Delta \bm{H} = \frac{1}{M_\mathrm{s} t} \left( \hat{\bm{m}}_0 \times \bm{p} + \beta \bm{p} \right)
= - \frac{1}{M_\mathrm{s} t} \left( \bm{p} \times \hat{\bm{m}}_0 + \beta' \bm{p} \right),
\end{aligned}
\end{equation}
where we defined 
$\beta' \equiv -\beta$.
If we adopt the form~(\ref{eq:sot:pcrossm}), Eq.~(\ref{eq:sot:aJbJ}) reads
\begin{equation}
\begin{aligned}
\label{eq:sot:aJbJ:pcrossm}
(\Delta H_x,& \Delta H_y, \Delta H_z)\\
&= (h_\mathrm{DL} \cos\theta_0, \beta' h_\mathrm{DL}, - h_\mathrm{DL} \sin\theta_0 \cos\varphi_0).
\end{aligned}
\end{equation}
Substituting Eq.~(\ref{eq:sot:aJbJ:pcrossm}) into the second line of Eq.~(\ref{eq:V2w:inp}), we obtain
\begin{equation}
\begin{aligned}
\label{eq:V2w:inp:aJbJ:pcrossm}
V_{2 \omega} &= \frac{1}{2} \Delta I a \left[ R_\mathrm{A} \frac{ h_\mathrm{DL} }{H - H_K} \cos\varphi_H \right.\\
& \ \ \ \ \ \ \ \ \ \ \ \ \ \ \ \ \ \left.  - 2 R_\mathrm{P} \frac{ \beta' h_\mathrm{DL} }{H} \cos\varphi_H \cos 2 \varphi_H \right].
\end{aligned}
\end{equation}
Equation~(\ref{eq:V2w:inp:aJbJ:pcrossm}) can be obtained by simply substituting $\beta' \equiv -\beta$ into Eq.~(\ref{eq:V2w:inp:aJbJ}).
Notice that the second term of the right hand side of Eq.~(\ref{eq:V2w:inp:aJbJ:pcrossm}) has a minus sign [compare this to Eq.~(\ref{eq:V2w:inp:aJbJ})]. 

The reason why the difference is important is because the sign of $\beta$ or $\beta'$ is often used to describe the sign of the field-like term. 
When $\beta$ is used, its sign is with respect to the damping-like term of the form $\hat{\bm{m}}_0 \times \bm{p}$, whereas the sign of $\beta'$ is in reference to the damping-like term of $\bm{p} \times \hat{\bm{m}}_0$.
Of course, if one defines the field-like term with respect to the coordinate axis, there is no ambiguity in the sign: see e.g. Eq.~(\ref{eq:sot}).
Experimentally, negative $\beta$ was reported in Ta/CoFeB\cite{kim2013nmat} and Pt/Co\cite{garello2013nnano}.
The sign can change depending on the material\cite{manchon2019rmp}, film thickness\cite{kim2013nmat}, growth condition\cite{pai2015prb} and even measurement temperature\cite{kim2014prb}.
Calculations have shown that negative $\beta$ can assist magnetization switching of perpendicular magnets\cite{taniguchi2015prb,legrand2015prap}.

\subsection{Sign of the effective anisotropy field}
Next, for the rotating-field harmonic Hall voltage measurements, the magnetic easy axis of the FM layer often points along the film plane. Under such circumstance, the effective out of plane magnetic anisotropy field $H_K$ is negative. 
It is thus common to define a positive quantity $H_K' \equiv - H_K > 0$. Substituting $H_K'$ into Eq.~(\ref{eq:V2w:inp:aJbJ:pcrossm}), we obtain
\begin{equation}
\begin{aligned}
\label{eq:V2w:inp:aJbJ:pcrossm:HK'}
V_{2 \omega} = \frac{1}{2} \Delta I a \left[ R_\mathrm{A} \frac{ h_\mathrm{DL} }{H + H_K'} \cos\varphi_H - 2 R_\mathrm{P} \frac{ \beta' h_\mathrm{DL} }{H} \cos\varphi_H \cos 2 \varphi_H \right].
\end{aligned}
\end{equation}

\subsection{Prefactor of $R_{2\omega}$}
Finally, there is a factor of two difference among some papers that uses the rotating-field harmonic Hall voltage measurements to extract the current-induced torque.
The difference likely originates from the conversion of $\sin^2 \omega t$ to $\cos 2\omega t$ [$\sin^2 \omega t = \frac{1}{2} \left(1 - \cos 2 \omega t \right)$] when deriving Eqs.~(\ref{eq:Vyx}) and (\ref{eq:V2w}).
The factor of $\frac{1}{2}$ in front of the right hand side of Eq.~(\ref{eq:V2w:inp:aJbJ:pcrossm:HK'}) is due to this conversion.
With regard to earlier publications, references~\cite{avci2014prb,lau2017jjap,chi2020sciadv,chi2021aplmater} missed this factor.

Substituting $a=1$ for the amplitudes of the anomalous and planar Hall resistances [see Eq.~(\ref{eq:Ryx})], and including the thermo-electric effects discussed in Refs~\cite{avci2014prb,roschewsky2019prb}, the harmonic Hall resistance $R_{2\omega}$ of the rotating-field measurements reads
\begin{widetext}
\begin{equation}
\begin{split}
\label{eq:R2w}
R_{2\omega} \equiv \frac{V_{2 \omega}}{\Delta I} = \frac{1}{2} \left( R_\mathrm{A} \frac{h_\mathrm{DL}}{H+H_K'} + \frac{r_\mathrm{ON} }{\Delta I } H + \frac{ V_\mathrm{c} }{\Delta I} \right) \cos\varphi 
- R_\mathrm{P} \frac{\beta' h_\mathrm{DL} + \eta H_\mathrm{Oe} }{H} \cos 2\varphi \cos\varphi,
\end{split}
\end{equation}
\end{widetext}
where the second term\cite{avci2014prb} on the right hand side represents contribution from the ordinary Nernst effect, the third term\cite{roschewsky2019prb} is the sum of contributions from the anomalous Nernst effect and the combined action of spin Seebeck effect and inverse spin Hall effect. 
Specifically, $r_\mathrm{ON}$ and $V_\mathrm{c}$ are defined as
\begin{equation}
\begin{aligned}
\label{eq:thermo}
r_\mathrm{ON} &= \mathcal{N} w \Delta T,\\
V_\mathrm{c} &= (\alpha_\mathrm{AN} + \alpha_\mathrm{SS}) w \Delta T,\\
\end{aligned}
\end{equation}
where $w$ is the width the channel of the Hall bar and $\Delta T$ is the temperature gradient along the film normal.
$\mathcal{N}$ is the ordinary Nernst coefficient, $\alpha_\mathrm{AN}$ is the anomalous Nernst coefficient, and $\alpha_\mathrm{SS}$ is a proportionality constant that represents the combined action of the spin Seebeck effect and inverse spin Hall effect in response to $\Delta T$.
In Eq.~(\ref{eq:R2w}), we included the Oersted field $H_\mathrm{Oe}$ in the last term on the right hand side as it can be numerically estimated\cite{hayashi2014prb}.
$H_\mathrm{Oe}$ includes information on $j_\mathrm{c}$: it changes sign when the current direction is reversed.
As described in Eq.~(\ref{eq:js:pol}), $\eta$ represents the film stacking: $\eta = 1$ for sub./NM/FM and $-1$ for sub./FM/NM.
Note that $\eta$ also appears in $h_\mathrm{DL}$: see Eq.~(\ref{eq:sot:aJbJ}).
With the thermo-electric effects included, we consider Eq.~(\ref{eq:R2w}) best represents the rotating-field harmonic Hall resistance for current-induced torque measurements.

Finally, the damping-like ($\xi_\mathrm{DL}$) and field-like ($\xi_\mathrm{FL}$) spin-torque efficiencies can be obtained from the following relations: 
\begin{equation}
\begin{aligned}
\label{eq:xi}
    \xi_\mathrm{DL} &= \frac{h_\mathrm{DL}}{j_\mathrm{c}} \frac{2 e}{\hbar} \frac{M_\mathrm{s} t}{\eta},\\
    \xi_\mathrm{FL} &= \frac{\beta' h_\mathrm{DL}}{j_\mathrm{c}} \frac{2 e}{\hbar} \frac{M_\mathrm{s} t}{\eta}.
\end{aligned}
\end{equation}
Note that here we divided the right hand side with $\eta$ so that the spin-torque efficiencies do not depend on the stacking order.

\section{Acknowledgements}
We thank Zhenchao Wen for insightful comments. 

\bibliography{spin}

\end{document}